\documentclass[]{jfm}
\usepackage{natbib}
\usepackage{amsmath}
\usepackage[dvips]{graphicx}
\usepackage{psfrag}

\newcommand{\sU}{\mathcal{U}}
\newcommand{\T}{\mathcal{T}}
\newcommand{\averzt}[1]{\langle #1 \rangle_{z,t}}

\begin{document}

\title[Streamwise-traveling waves for drag reduction]{Streamwise-traveling waves \\ of spanwise wall velocity \\ for turbulent drag reduction}

\author[M. Quadrio, P. Ricco \& C. Viotti]{
M\ls A\ls U\ls R\ls I\ls Z\ls I\ls O\ls \ns Q\ls U\ls A\ls D\ls R\ls I\ls O$^1$,
P\ls I\ls E\ls R\ls R\ls E\ls\ns R\ls I\ls C\ls C\ls O$^2$\ls
\and
C\ls L\ls A\ls U\ls D\ls I\ls O\ns V\ls I\ls O\ls T\ls T\ls I$^3$ 
}

\affiliation{
$^1$Politecnico di Milano, 20156 Milano, Italy \\
$^2$King's College London, WC2R 2LS London, United Kingdom \\
$^3$The University of North Carolina, 27516 Chapel Hill, NC, USA.
}

\maketitle

\begin{abstract}

Waves of spanwise velocity imposed at the walls of a plane turbulent channel flow are studied by Direct Numerical Simulations. We consider sinusoidal waves of spanwise velocity which vary in time and are modulated in space along the streamwise direction. The phase speed may be null, positive or negative, so that the waves may be either stationary or traveling forward or backward in the direction of the mean flow. Such a forcing includes as particular cases two known techniques for reducing friction drag: the oscillating wall technique (a traveling wave with infinite phase speed) and the recently proposed steady distribution of spanwise velocity (a wave with zero phase speed).

The traveling waves alter the friction drag significantly. Waves which slowly travel forward produce a large reduction of drag, that can relaminarize the flow at low values of the Reynolds number. Faster waves yield a totally different outcome, i.e. drag increase. Even faster waves produce a drag reduction effect again. Backward-traveling waves instead lead to drag reduction at any speed.

The traveling waves, when they reduce drag, operate in similar fashion to the oscillating wall, with an improved energetic efficiency. Drag increase is observed when the waves travel at a speed comparable with that of the convecting near-wall turbulence structures. A diagram illustrating the different flow behaviors is presented.

\end{abstract}

\section{Introduction}
\label{sec:introduction}

Reducing the friction drag in turbulent wall flows has seen a growing interest in recent years, owing to the awareness for high energy consumption and to the need of decreasing pollutant emissions into the atmosphere. Drag reduction techniques are usually grouped \citep{gadelhak-2000} into active (requiring external energy to be fed into the system) and passive techniques. Another useful classification is between closed-loop strategies (requiring a control law with feedback) and simpler open-loop techniques: the review by \cite{kim-2003} discusses the former group, whereas the present work is concerned with the latter.

One interesting open-loop technique, which is conceptually simple and with the potential for a positive energetic budget, is the cyclic spanwise movement of the wall. In this active technique, first introduced by \cite{jung-mangiavacchi-akhavan-1992}, the wall moves sinusoidally with period $T$ according to
\begin{equation}
\label{eq:t}
w_w(t) = A \sin \left( \omega t \right),
\end{equation}
where $w_w$ is the spanwise ($z$) component of the velocity vector at the wall, $t$ is time, $A$ is the oscillation amplitude and $\omega = 2 \pi / T$ is the oscillation frequency. The main features of the oscillating-wall technique \citep{ricco-quadrio-2008} are the existence of an optimal frequency $\omega_{opt}$ for drag reduction at fixed $A$, and a significant maximum drag reduction, which can be as high as 45\% when $A$ is comparable with the flow centreline velocity. At lower $A$, the energy balance between the power saved thanks to the oscillation and the power spent to activate the wall motion against the viscous stresses may be positive at about 7\% when $\omega=\omega_{opt}$. 

In a recent work, \cite{quadrio-viotti-luchini-2007} \citep[see also][]{viotti-quadrio-luchini-2008} have addressed one of the drawbacks of the oscillating-wall technique, i.e. its unsteady nature. They converted the time-dependent forcing (\ref{eq:t}) into its stationary counterpart of a spatial sinusoidal distribution of spanwise velocity over a wavelength $\lambda_x$:
\begin{equation}
\label{eq:x}
w_w(x) = A \sin \left( \kappa_x x \right),
\end{equation}
where $x$ is the streamwise coordinate and $\kappa_x = 2 \pi / \lambda_x$ is the streamwise wavenumber. The time dependence of Eq.(\ref{eq:t}) is converted into a relatively unsteady interaction between the steady wall motion and the convecting near-wall turbulence. Such conversion is only apparently similar to the Taylor's frozen-turbulence hypothesis, for which the velocity scale is the mean streamwise velocity. Indeed, the mean velocity is zero at the wall, while the turbulence keeps a strongly convective character \citep{kreplin-eckelmann-1979, kim-hussain-1993}, with elongated space-time correlations \citep{quadrio-luchini-2003}. The velocity scale for space-time conversion is therefore the convection velocity $\sU$ of the turbulent fluctuations, which almost coincides with the mean velocity along with the wall-normal span of the channel, except for a thin near-wall layer, say $y^+ < 10$, where it is constant at $\sU_w^+ \approx 10$. (Hereinafter, the $+$ superscript indicates scaling by the friction velocity $u_\tau$ of the reference flow and the kinematic viscosity $\nu$ of the fluid.) \cite{viotti-quadrio-luchini-2008} have shown that the optimal frequency $\omega_{opt}$ for (\ref{eq:t}) translates into an optimal wavenumber $\kappa_{x,opt}$ for (\ref{eq:x}) through the relation
\begin{equation}
\label{eq:conversion-x-t}
\kappa_{x,opt} = \frac{\omega_{opt}}{\sU_w}.
\end{equation}

The present paper addresses the extension of (\ref{eq:t}) and (\ref{eq:x}) to the space-time case
\begin{equation}
\label{eq:x-t}
w_w(x,t) = A \sin \left( \kappa_x x - \omega t \right),
\end{equation}
where $\kappa_x$ and $\omega$ may be both different from zero at the same time. In (\ref{eq:x-t}), the wave of spanwise velocity moves (backward or forward) in the streamwise direction, and its phase speed is
\begin{equation}
 c = \frac{\omega}{\kappa_x}.
\label{eq:c}
\end{equation}
The temporal (\ref{eq:t}) and steady (\ref{eq:x}) oscillations correspond respectively to a wave traveling at infinite speed ($\kappa_x=0$), and to a stationary wave ($\omega=0$). The effect of the wall motion is already known on the axes of the $\omega - \kappa_x$ plane, while it is unexplored in the rest of the plane. The main question addressed in this paper is therefore: what is the response of the turbulent flow in the whole $\omega - \kappa_x$ plane?

Before presenting our results, it may be useful to mention the drag reduction technique by \cite{du-karniadakis-2000} and \cite{du-symeonidis-karniadakis-2002}. They studied the following spanwise-oriented volume forcing:
\begin{equation}
\label{eq:z-t}
 f_z(z,t) = F \ \mathrm{e}^{-y / \Delta} \sin \left( \kappa_z z - \omega t \right),
\end{equation}
where $\Delta$ is the distance up to which the forcing diffuses from the wall and $\kappa_z$ is the spanwise wavenumber. In a turbulent channel flow at $Re_\tau=150$, they obtained about 30\% drag reduction and observed a substantial disruption of the near-wall flow structures. The volume force in Eq.(\ref{eq:z-t}) consists of {\em spanwise}-traveling waves acting along the spanwise direction, while here we consider a wall-based motion of spanwise velocity waves traveling along the {\em streamwise} direction.

In the present paper, the effects of imposing condition (\ref{eq:x-t}) are investigated by Direct Numerical Simulations (DNS). While purely spatial and purely temporal oscillations have been shown by \cite{viotti-quadrio-luchini-2008} to be largely analogous to each other, the flow response to the traveling waves is found to be interestingly complex when examined as a function of $\omega$ and $\kappa_x$. Both drag reduction (DR) and drag increase (DI) may occur. The maximum DR resides in a non-obvious region of the $\omega - \kappa_x$ plane, which largely corresponds to where the global energy balance is positive, whereas the maximum DI occurs where the phase speed $c$ is comparable with the convection velocity $\sU_w$. 

The structure of the paper is as follows. In \S\ref{sec:method}, the numerical procedures are outlined. Section \S\ref{sec:results} presents the drag modification and the global energy budget in the $\omega - \kappa_x$ plane, and discusses the effect of forcing amplitude and Reynolds number on the maximum DR. Selected flow statistics are also described. In \S\ref{sec:discussion}, the analogy between the oscillating wall and the traveling waves is discussed, and a diagram offers a comprehensive view on the modified turbulent flow. A brief summary is given in \S\ref{sec:summary}.

\section{Method}
\label{sec:method}

A large number of DNS has been carried out to examine the response of a turbulent channel flow to the wall forcing (\ref{eq:x-t}). The traveling waves are applied at both walls and move in phase. From a numerical perspective, Eq.(\ref{eq:x-t}) is a non-homogeneous, time-dependent wall boundary condition for the spanwise component of the velocity vector. The size of the computational domain, the requirements in terms of the spatio-temporal resolution and the length of the integration interval render these DNS very demanding. The efficiency of the parallel DNS code and the availability of appropriate computational resources are key aspects of this research effort.

The code and the architecture of the computing system have been described by \cite{luchini-quadrio-2006}. The code is a mixed-discretization parallel solver of the incompressible Navier--Stokes equations, based on Fourier expansions in the homogeneous directions and high-order, explicit compact finite-difference schemes in the wall-normal direction. Its primary design target is to achieve high performance while using off-the-shelf hardware. Most of the calculations have been run on a computing system available in dedicated mode at the Universit\`a di Salerno. The system is composed of 268 AMD Opteron CPUs, grouped in 134 computing nodes, and connected with an {\em ad hoc} topology through Gigabit Ethernet cards. Each node is equipped with its own hard disk and RAM. The system possesses 0.28 TB of RAM and 40 TB of storage space, with a peak computing power of 2.6 TFlop/s.

The computations are integrated forward in time, starting from the same initial condition of a fully developed channel flow without waves. The flow rate is constant and the bulk velocity is $U_b = 2/3 U_P$, where $U_P$ is the centreline velocity of a laminar Poiseuille flow with the same flow rate. About 250 simulations have been run with a wave amplitude $A = 0.5 U_P$ (which corresponds to $A^+=12$) at a Reynolds number $Re=U_P h/\nu=4760$ (which corresponds to $Re_\tau = h u_\tau / \nu = 200$ based on $u_\tau$ and the channel half width $h$). A few cases have also been run to explore how the maximum DR is sensitive to changes in $A$ or $Re$: $A/U_P$ has been varied in the range $0.085-1.26$, and two values of $Re=2175$ and $Re=10500$ (corresponding to $Re_\tau=100$ and $Re_\tau=400$, respectively) have been considered.

\begin{figure}
\vspace{1cm}
\centering
\psfrag{F}{Mean streamwise flow}
\psfrag{x}{$x$}
\psfrag{y}{$y$}
\psfrag{z}{$z$}
\psfrag{Lx}{$L_x$}
\psfrag{Ly}{$L_y=2h$}
\psfrag{Lz}{$L_z$}
\psfrag{W}{$w = A \sin \left(\kappa_x x - \omega t \right)$}
\psfrag{U}{$c$}
\psfrag{l}{$\lambda_x$}
\centering
\includegraphics[width=0.85\textwidth]{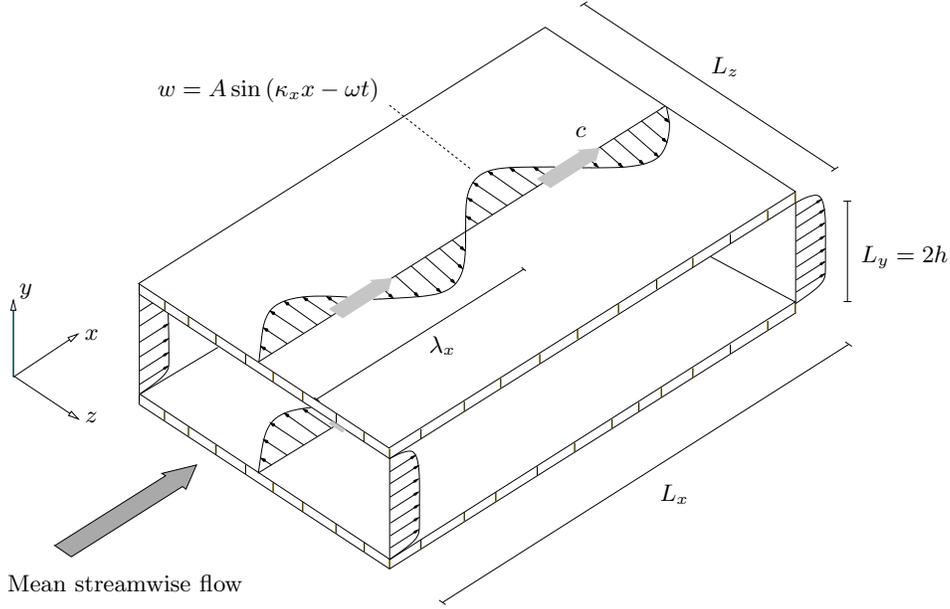}
\caption{Schematic of the system for turbulent channel flow with wall traveling waves. $\lambda_x$ is the streamwise wavelength and $c$ is the phase speed of the waves (traveling forward in this sketch). $L_x$, $L_y$ and $L_z$ are the dimensions of the computational domain in the streamwise, wall-normal and spanwise directions, respectively.}
\label{fig:traveling-sketch}
\end{figure}

The computational domain, a schematic of which is presented in figure \ref{fig:traveling-sketch}, has a streamwise length of $L_x = 6 \pi h$ and a spanwise width of $L_z = 3 \pi h$. At $Re=4760$, a Fourier expansion of $320 \times 320$ modes is used in the streamwise and spanwise directions, and $160$ points discretize the wall-normal direction, for a total of $3.3 \cdot 10^7$ spatial degrees of freedom. The streamwise and spanwise resolutions are $\Delta x^+ = 11.8$ and $\Delta z^+ = 5.9$, and the wall-normal resolution varies from $\Delta y^+_{min} = 0.7$ at the wall to $\Delta y^+_{max} = 4.1$ at the channel centreline. The time step is $\Delta t^+ = 0.16$, and the total integration time is 1000  $h / U_P$ (i.e. 8400 viscous time units). A few cases (first row of data at $\kappa_x h = 0.167$ in figure \ref{fig:k-omega-DR}) have $L_x=12 \pi h$, and use 640 streamwise Fourier modes to keep the resolution unchanged. A few other cases (with extremely large oscillation periods) have an averaging time of 2000  $h / U_P$. For the simulations at different $Re$, the size of the computational domain has been kept constant in outer units, and the spatial resolution has been kept constant in viscous units. The effect of the discretization parameters on the friction drag has been assessed by preliminary checks. In agreement with \cite{quadrio-ricco-2004}, the accuracy in measuring the friction coefficient is better than 1\%, and can be further improved by extending the total averaging time. 

The main objective is to investigate how the forcing (\ref{eq:x-t}) alters the friction coefficient, defined as
\[
C_f = \frac{2 \tau_w}{\rho U_b^2},
\]
where $\tau_w$ is the mean wall-shear stress and $\rho$ is the density of the fluid.
Changes of $C_f$ are quantified in terms of percentage of $C_{f,0}$, the friction coefficient of the reference turbulent flow, and coincide with the percentage saving of the energy required to drive the fluid along the $x$ direction at fixed flow rate. For the reference case at $Re=4760$, $C_{f,0} = 7.94 \cdot 10^{-3}$, which is in very good agreement with the correlation reported by \cite{pope-2000} at page 279. As in \cite{quadrio-ricco-2004}, $C_f$ is evaluated by space- and time-averaging, after discarding the initial transient during which the flow adapts to the new regime. The length of the transient is estimated by direct observation of the time history of the space-averaged friction, and is typically of the order of 100-200$h/U_p$. Unless otherwise indicated, quantities reported hereinafter are scaled by  $h$ and $U_P$.

\section{Results}
\label{sec:results}

\subsection{Effect on friction drag}
\label{sec:drag}

\begin{figure}
\centering
\psfrag{k}{$\kappa_x$}
\psfrag{w}{$\omega$}
\includegraphics[angle=90,width=1.1\textwidth]{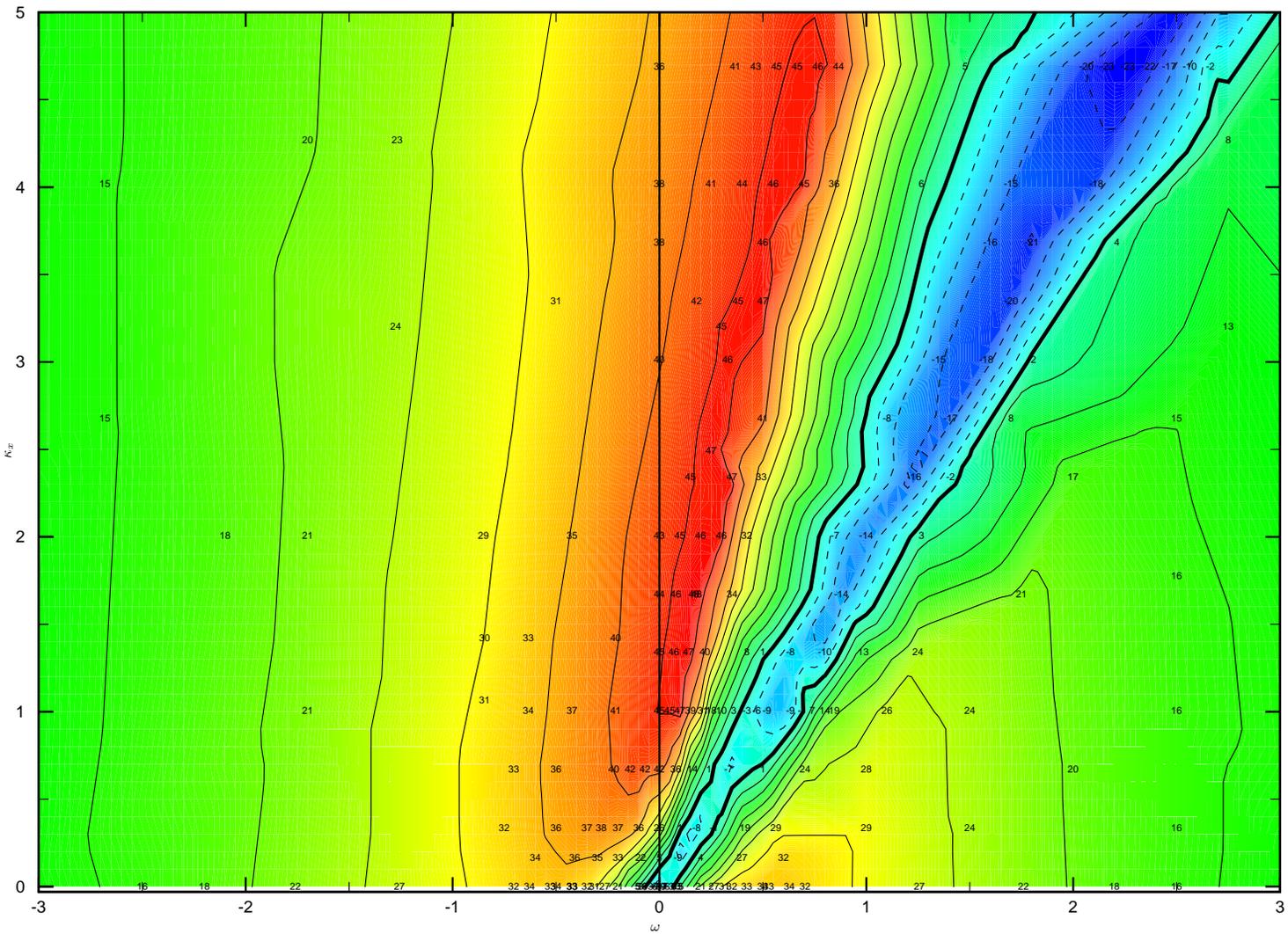}
\caption{Map of friction drag reduction (percentage) in the $\omega - \kappa_x$ plane for $A=0.5$ and $Re=4760$. Contours are spaced by 5\% intervals, loci of zero DR are indicated by thick lines and negative values are represented by dashed lines. The numbers indicate percentage drag reduction at measured points.}
\label{fig:k-omega-DR}
\end{figure}

Figure \ref{fig:k-omega-DR} describes the effects of the traveling waves in terms of the percentage change in friction drag as a function of $\omega$ and $\kappa_x$ for $A=0.5$ and $Re=4760$. Only the upper half of the $\omega - \kappa_x$ plane is shown, since (\ref{eq:x-t}) yields symmetric results upon exchanging the pair $\omega,\kappa_x$ with $-\omega,-\kappa_x$. (This has been explicitly verified for a few points.) The phase speed $c$ is interpreted graphically as the inverse of the slope of straight lines passing through the origin. In the first quadrant, the waves move forward in the direction of the mean flow ($c>0$), while the second quadrant corresponds to backward-traveling waves ($c<0$).

Each point in figure \ref{fig:k-omega-DR} is the result of one simulation. The contours assist in visualizing the general behaviour of the changes in drag; they are computed after a linear interpolation of the irregularly scattered points on a regular mesh with grid sizes of $\Delta \omega = 0.05$ and $\Delta \kappa_x = 0.1$. The presence of the numerical values, rounded to the nearest integer, allows verifying that the contours do not misrepresent the data. Several points with large $\omega$ and/or $\kappa_x$ do not appear, but they are however useful to obtain the correct trend of the contours near the boundaries.

On the horizontal $\omega$ axis, Eq.(\ref{eq:x-t}) reduces to the oscillating-wall case (\ref{eq:t}): our data agree with those by \cite{quadrio-ricco-2004}, who used the same $Re$ and slightly different discretization parameters. The symmetry mentioned above requires points at $\omega$ and $-\omega$ to have the same friction drag: this is enforced by plotting every data point twice, at $+\omega$ and $-\omega$. On the vertical $\kappa_x$ axis, Eq.(\ref{eq:x-t}) reduces to the steady wave case (\ref{eq:x}). Some of these points are taken from \cite{viotti-quadrio-luchini-2008}, who employed the same $Re$ and discretization parameters. As discussed there, DR behaves very similarly on the two axes, once frequency is converted into wavenumber through Eq.(\ref{eq:conversion-x-t}), reaching its maximum at $\omega = \omega_{opt} \approx 0.5$, and at $\kappa_x = \kappa_{x,opt} \approx 1$, with $\omega_{opt}$ and $\kappa_{x,opt}$ related by $\sU_w=0.5$.

Off the axes, the flow response reveals an unexpectedly rich behaviour. The contours show that local maxima and local minima form two elongated narrow regions. In the red region, whose crest identifies a curve which does not cross the origin, the flow presents an intense drag reduction. The maximum is about 48\%, which is higher than 34\% for the oscillating wall at $\omega_{opt}$ and 45\% for the steady wave at $\kappa_{x,opt}$. A wide range of $\kappa_x$ exists where DR is 47\%--48\% for a non-zero phase speed. The crest crosses the vertical axis at $\kappa_x = \kappa_{x,opt}$ and intersects the $\omega$ axis at $\omega = -\omega_{opt}$. For $\kappa_x > \kappa_{x,opt}$, this region largely corresponds to slow forward-traveling waves ($c \approx 0.15$), while for $\kappa_x < \kappa_{x,opt}$ it relates to backward-traveling waves. In the blue region, the drag increases significantly, up to 23\%. The maximum increase corresponds to a straight line crossing the origin, and is therefore associated with a constant phase speed, $c = 0.5$. The narrow, cone-shaped DI region is confined between the two thick zero-DR lines, corresponding to $c \approx 0.35$ and $c \approx 0.6$.
The marked asymmetry between the first and second quadrant vanishes at high $| \omega |$: the contour lines tend to become vertical, suggesting that in this limit the effect of the streamwise modulation becomes negligible and $\omega$ alone dictates the drag reduction level.

\begin{figure}
\centering
\psfrag{T}{$T$}
\psfrag{D}{$DR$(\%)}
\includegraphics[width=\textwidth]{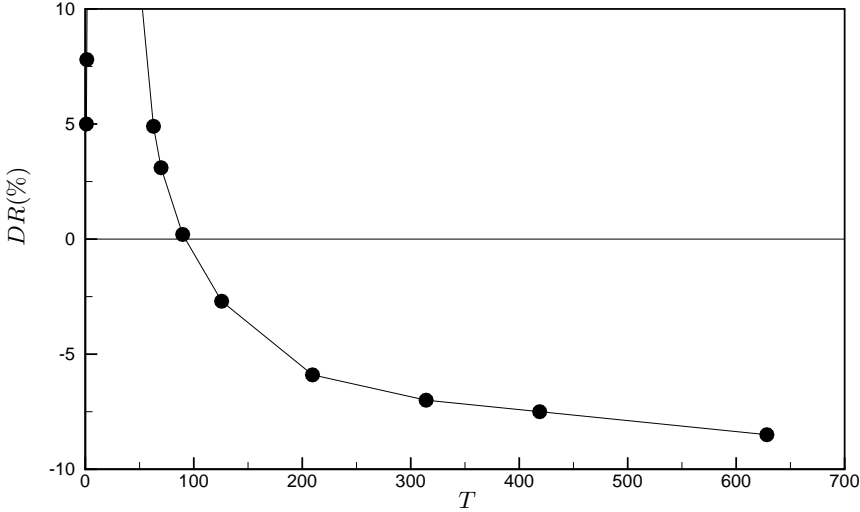}
\caption{Friction drag reduction (percentage) as a function of $T=2\pi/\omega$ for $A=0.5$, $Re=4760$ and $\kappa_x=0$. The leftmost points are at $T=1$ and $T=1.4$, and points with $2<T<50$ are out of scale.}
\label{fig:infrared}
\end{figure}

Careful observation of the region near the origin of the $\omega - \kappa_x$ plane reveals a finite DI for the oscillating wall at very small frequencies. This is shown in figure \ref{fig:infrared}, where $T=2 \pi / \omega$ is used as independent variable. \cite{nikitin-2000} was the first to surmise the existence of what he called an infrared DI, based on an under-resolved DNS at $T > 120$.  \cite{quadrio-ricco-2004} tested the oscillating-wall technique in a plane channel up to $T \approx 90$ and were unable to confirm the infrared DI (figure \ref{fig:infrared} indeed confirms that no drag increase is observed for $T<100$). Owing to the computational cost, we did not explore this issue further.

\subsection{Power budget}

\begin{figure}
\centering
\psfrag{k}{$\kappa_x$}
\psfrag{w}{$\omega$}
\includegraphics[width=\textwidth]{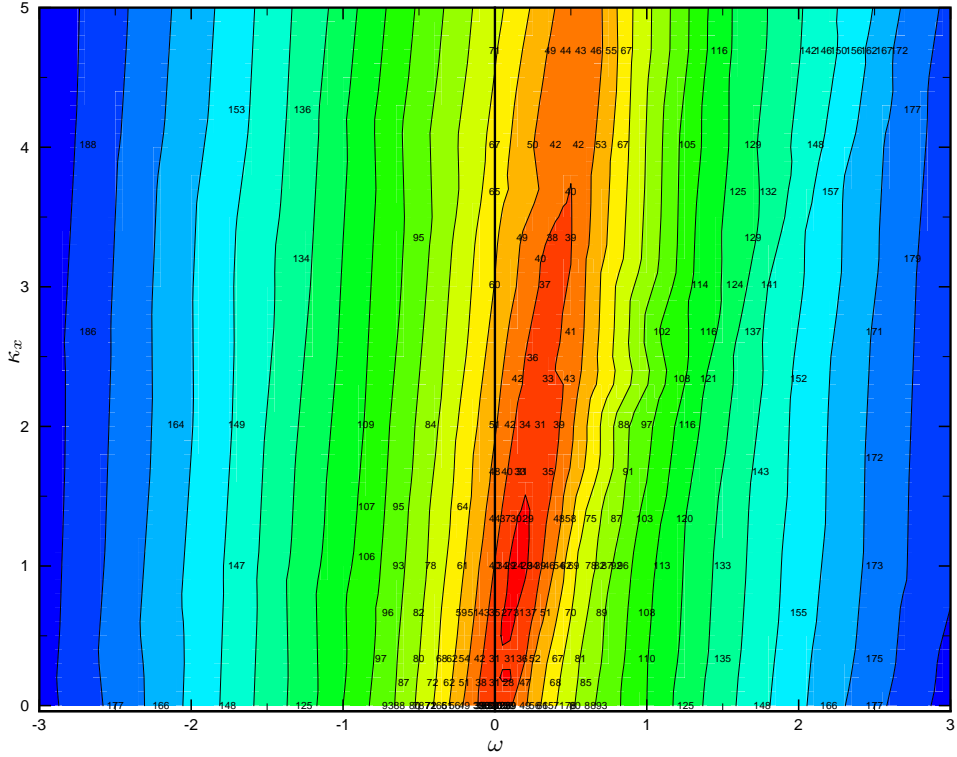}
\caption{Map of the power $P_{sp}$(\%) spent for the traveling waves in the $\omega - \kappa_x$ plane for $A=0.5$ and $Re=4760$. Contours are spaced by 10\% intervals and the minimum is 30\%. Here and in the following figure, the percentage is relative to the reference friction power required to drive the flow along the streamwise direction. The numbers indicate $P_{sp}$(\%) at measured points.}
\label{fig:k-omega-Preq}
\end{figure}

The efficiency of the traveling waves in affecting the friction drag is studied by computing the amount of external energy spent to enforce the control action. The quantification of this energy obviously implies the knowledge of the actuation device. However, the power that an ideal device dissipates against the viscous stresses provides an upper bound for the performance of the control strategy. We express this power as percentage of the power spent to drive the fluid along the streamwise direction in the fixed-wall configuration:
\[
P_{sp}(\%) = \frac{100}{(\mbox{d}U/\mbox{d}y|_0) U_b}
             \frac{1}{L_x L_z T}
\int_0^{L_x} \int_0^{L_z} \int_0^T \left. w_w \frac{\partial w}{\partial y} \right|_{y=0}
\mbox{d} t \mbox{d} z \mbox{d} x,
\]
where $\mbox{d}U/\mbox{d}y|_0$ is the gradient of the mean streamwise velocity at the wall in the reference case. $P_{sp}$(\%) is reported in figure \ref{fig:k-omega-Preq} as a function of $\omega$ and $\kappa_x$. Much less energy is required to impose a standing wave than to generate uniform oscillations, if cases related through Eq.(\ref{eq:conversion-x-t}) are compared. However, neither the oscillating wall nor the standing waves are associated with the pair which offers the minimum energetic expense of $P_{sp}(\%)=23.5$, i.e. $\omega \approx 0.15, \kappa_x \approx 1$.

\begin{figure}
\centering
\psfrag{k}{$\kappa_x$}
\psfrag{w}{$\omega$}
\includegraphics[width=\textwidth]{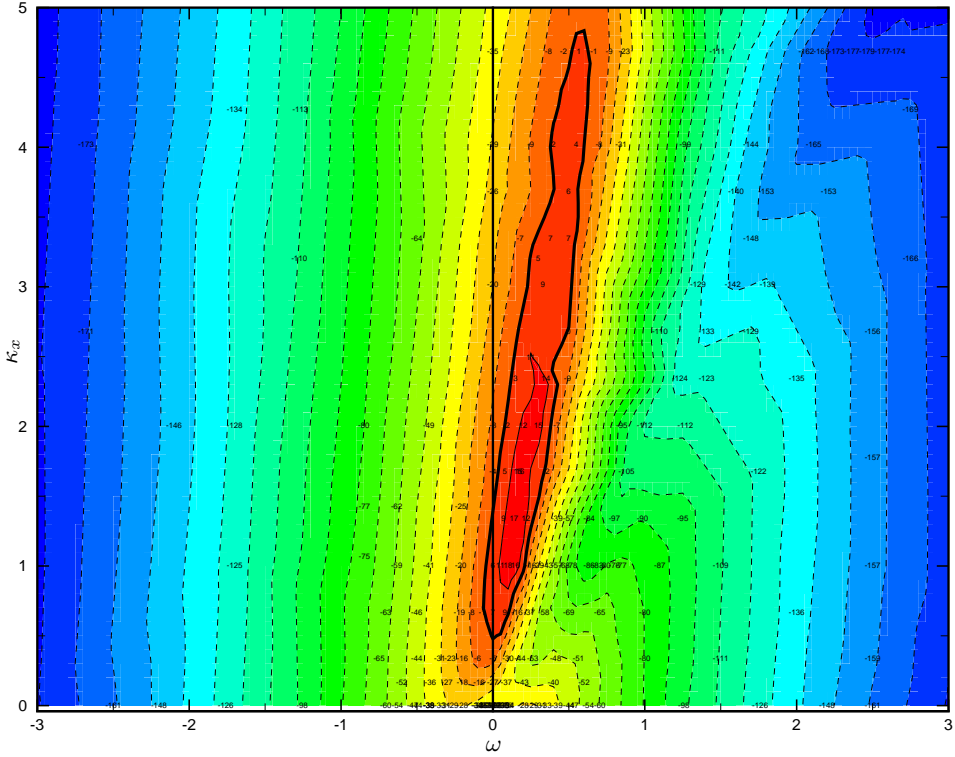}
\caption{Map of net power saving $P_{net}$(\%) in the $\omega - \kappa_x$ plane for $A=0.5$  and $Re=4760$. Contours are spaced by 10\% intervals. Locus of zero $P_{net}$(\%) is indicated by thick curve, solid lines denote positive balance and negative values are represented by dashed lines. The numbers indicate $P_{net}$(\%) at measured points.}
\label{fig:k-omega-Pnet}
\end{figure}

Even more important is to quantify the net energy saving, defined as the difference between the percentage power saved to drive the fluid along the streamwise direction (which coincides with the drag reduction at constant $U_b$) and the percentage power spent defined above, namely
\[
P_{net}(\%) = DR(\%) - P_{sp}(\%).
\]
For several active open-loop techniques, $P_{net}$(\%) is negative. Some methods, like the oscillating wall \citep{baron-quadrio-1996}, may offer a positive net budget of the order of a few percent. The traveling waves, on the other hand, guarantee much larger benefits because of the remarkable occurrence that the region of minimum $P_{sp}$(\%) largely coincides with the region of maximum $DR$(\%). Figure \ref{fig:k-omega-Pnet} shows that $P_{net}$(\%) is positive mainly for slow forward-traveling waves, and that a 18\% of maximum net is measured for $\omega \approx 0.15, \kappa_x \approx 1$ (which is also the optimum pair for minimum $P_{sp}$). This benefit is noteworthy when compared with the negative net balance of the oscillating wall at this value of $A$, and the small net gain of 5\% obtained by the steady waves. It must be recalled that we have discussed a simple estimate, which cannot account for the nature of the real actuation system. However, such a high positive net power budget might well accommodate a real system with efficiency less than unity.

\subsection{Effect of $A$ and $Re$ on maximum drag reduction}
\label{sec:effects-A-Re}

\begin{figure}
\centering
\psfrag{D}{$DR$(\%)}
\psfrag{N}{$P_{net}$(\%)}
\psfrag{A}{$A$}
\includegraphics[width=\textwidth]{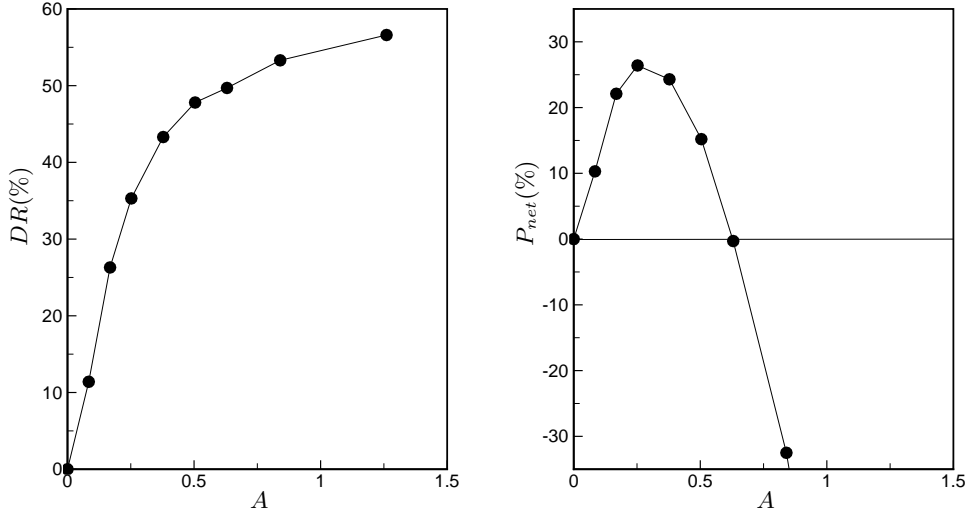}
\caption{Percentage drag reduction (left) and net power saving (right) as functions of forcing amplitude $A$. The forcing conditions are $\omega=0.16,\kappa_x=1.66$, which produce the maximum drag reduction at $A=0.5$ and $Re=4760$.}
\label{fig:amplitude}
\end{figure}

The sensitiveness of the maximum DR to the parameters $A$ and $Re$, kept fixed in the main parametric study, is addressed through a few additional simulations.
Figure \ref{fig:amplitude} (left) shows that DR increases monotonically with $A$ and reaches nearly 60\% when $A$ is slightly larger than the flow centreline velocity. In the plot on the right, $P_{net}$ grows significantly at lower amplitude, which is expected as the power spent to impose the traveling waves decreases quadratically with $A$, and is larger than 26\% for $A=0.25$. Both results are consistent with (but quantitatively much better than) the oscillating-wall scenario, studied by \cite{quadrio-ricco-2004}.

The maximum DR at $Re=4760$ depends weakly on the Reynolds number. We have observed full relaminarization of the flow when $Re=2175$ ($Re_\tau=100$), and a slight decrease from 48\% to 42\% as $Re$ increases to 10500 ($Re_\tau=400$). This finding is consistent with experimental \citep{choi-graham-1998,ricco-wu-2004} and DNS results \citep{ricco-quadrio-2008} available for the oscillating-wall case. Following these works, we have assumed that the values of $\omega$ and $\kappa_x$ that guarantee the maximum DR scale in wall units of the reference flow.

\subsection{Turbulence statistics}
\label{sec:statistics}

\begin{figure}
\centering
\includegraphics[width=\textwidth]{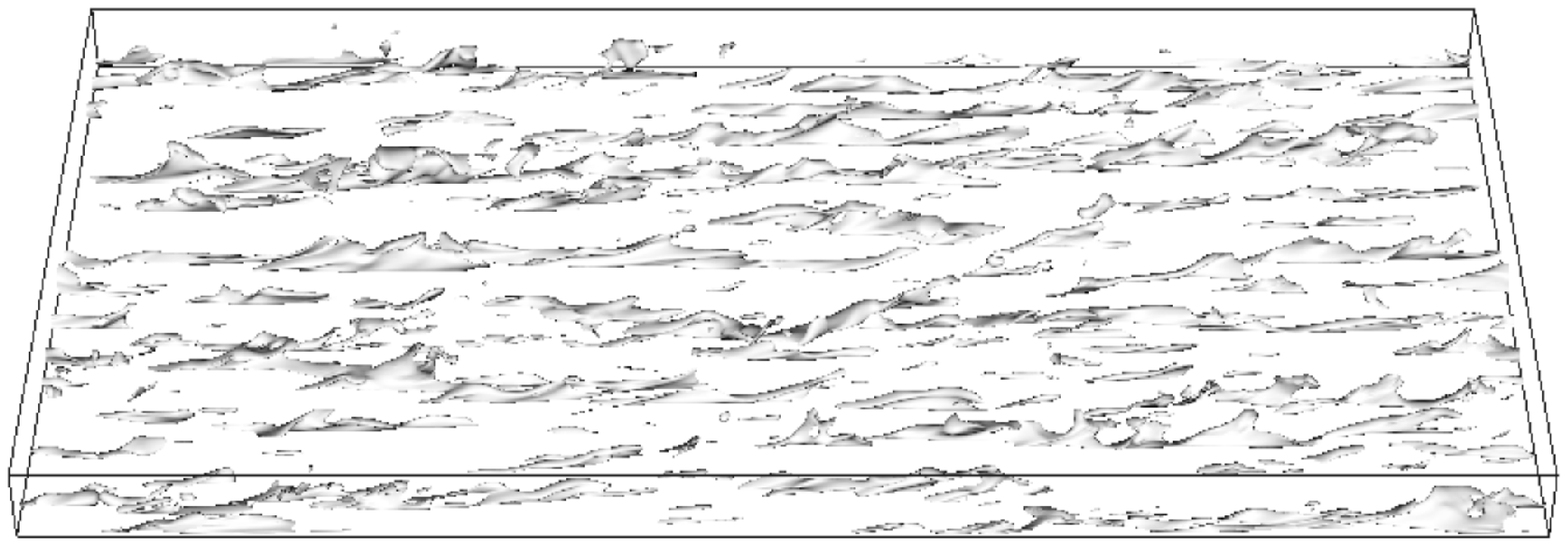}
\includegraphics[width=\textwidth]{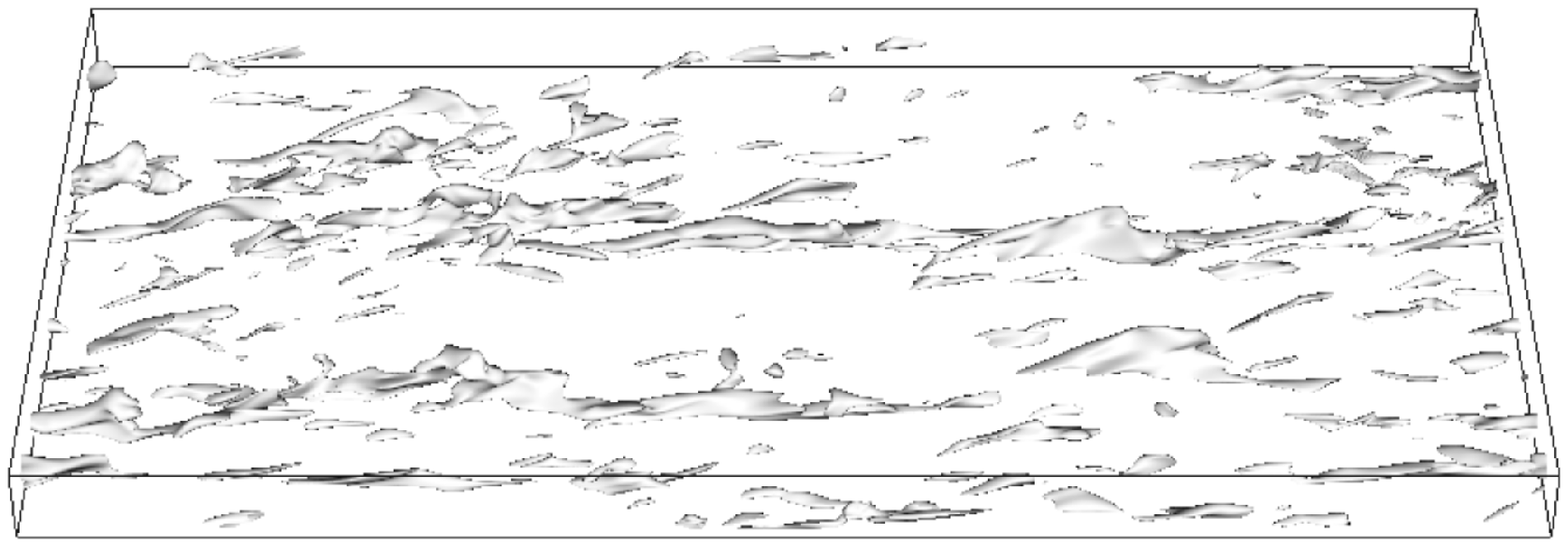}
\includegraphics[width=\textwidth]{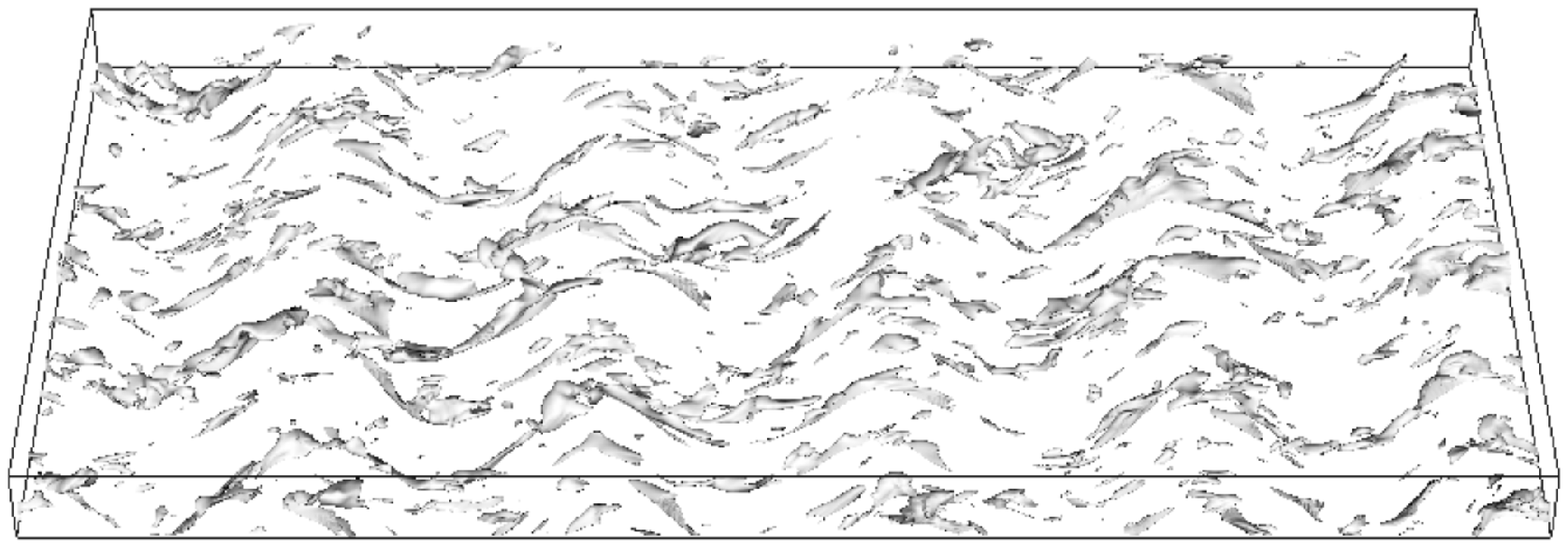}
\caption{Snapshots of instantaneous flow fields for the reference flow (top), a case with DR (middle, $\kappa_x=1.66$ and $\omega=0.16$) and a case with DI (bottom, $\kappa_x=1.66$ and $\omega=0.88$). Isosurfaces for the quantity $q=\mbox{sgn}(u)\sqrt{u^2+w^2}$ at the value $q^+=-4$. The value $q^+=-4$ is computed with the actual friction velocity of the drag-modified flow. The flow is from left to right and only the bottom half of the channel is shown. For both cases $A=0.5$ and $Re=4760$.}
\label{fig:snapshot}
\end{figure}

In this section, we study the turbulence statistics to understand how the flow is modified by the traveling waves. We first take a preliminary look at instantaneous snapshots of the flow field in figure \ref{fig:snapshot}. Two different cases are compared to a snapshot of the reference flow, shown in the top plot: one with $\omega=0.16,\kappa_x=1.66$ and the other at $\omega=0.88,\kappa_x=1.66$. These two cases do not lie far from each other in the $\omega - \kappa_x$ plane, but yield totally different outcomes, i.e. a large DR of 48\% for the former (middle plot) and a large DI of 14\% for the latter (bottom plot). Both cases refer to forward-traveling waves, with $c=0.1$ and $c=0.53$, respectively. 

Isosurfaces for negative values of the quantity $q=\mbox{sgn}(u)\sqrt{u^2+w^2}$ are shown in figure \ref{fig:snapshot} to visualize the low-speed streaks of the near-wall flow even in the present case where the local streamwise direction may deviate from the $x$ axis. The value $q=-0.168$ ($q^+=-4$) is chosen for the reference flow, that indeed shows the usual pattern of near-wall elongated, streamwise-aligned regions. In the plots for the DR and DI cases, the value of $q$ is changed so as to always yield $q^+=-4$ when computed with the actual friction velocity of the flow. This compensates for the largely different values of $Re_\tau$: had the nominal value of $u_\tau$ been used, the streaky structures would have disappeared in the DR case. When DR takes place, what remains after accounting for the reduced $Re_\tau$ are elongated structures, aligned with the $x$ direction, of perhaps larger size and reduced streamwise meandering. In sharp contrast, in the DI case the same structures present an evident streamwise modulation, with a streamwise wavelength corresponding to the wavelength of the wall wave, which is a marked structural difference with the reference flow. In this case the most evident action of the forcing is to cyclically change the direction of the flow, by tilting the low-speed streaks in accordance with the local spanwise velocity component at the wall.

\begin{figure}
\centering
\psfrag{U}{$U$}
\psfrag{y}{$y$}
\includegraphics[width=0.8\textwidth]{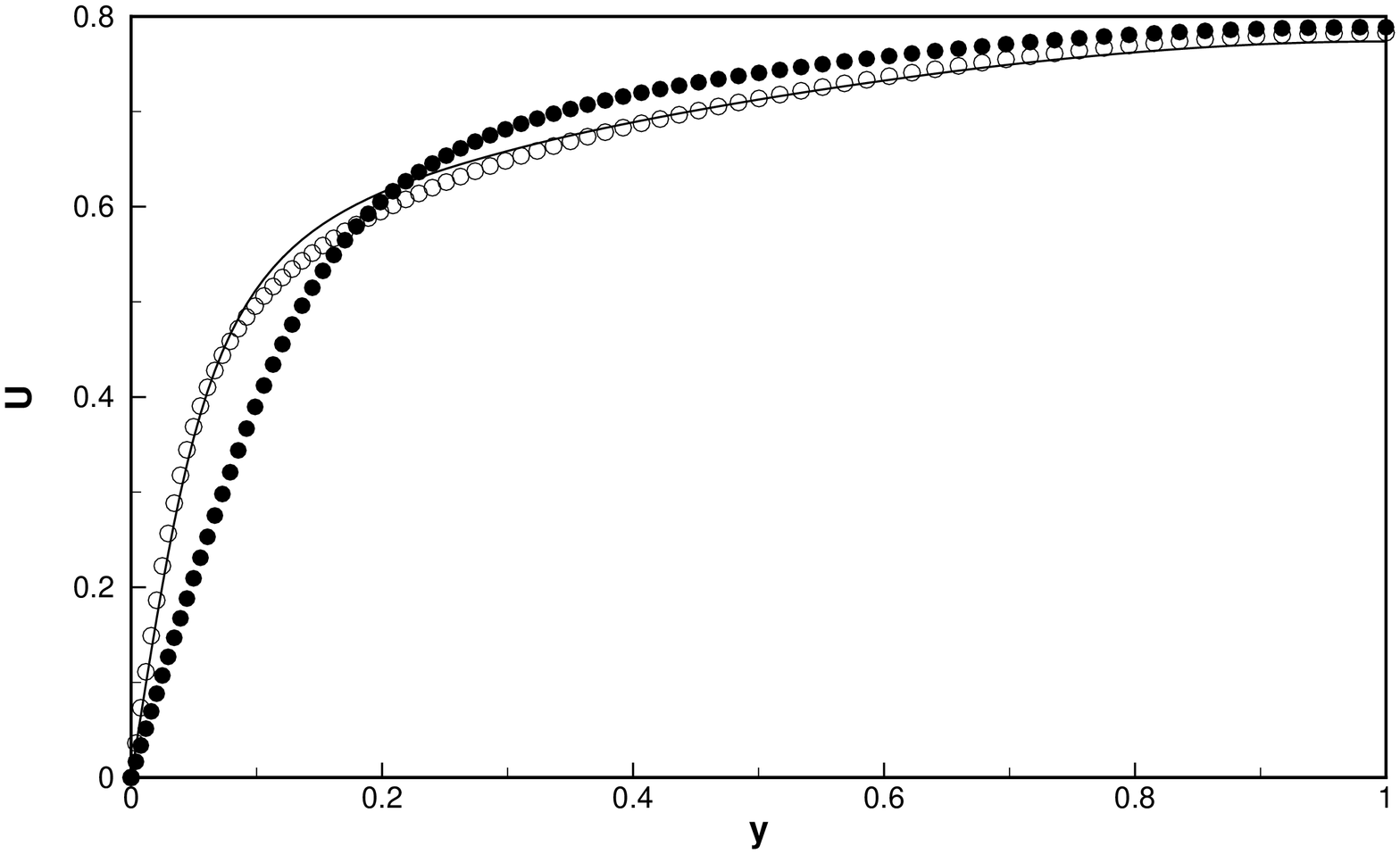}
\caption{Profiles of mean streamwise velocity $U$, after averaging over time, $x$ and $z$. The solid line is the reference case, the empty circles are the DI case $(\omega=0.88,\kappa_x=1.66)$, and the filled circles are the DR case $(\omega=0.16,\kappa_x=1.66)$. For both cases $A=0.5$ and $Re=4760$.}
\label{fig:u-mean}
\end{figure}

In figure \ref{fig:u-mean}, the profiles of the mean streamwise velocity component, averaged over time, $x$ and $z$, are presented as functions of the wall-normal coordinate. They are indicated by the solid line (reference case), filled (DR) and empty (DI) circles. The effect of the wall motion diffuses all along the extent of the channel and the large near-wall change of the velocity gradient is evident when DR occurs.

\begin{figure}
\centering
\psfrag{x}{$\xi$}
\psfrag{y}{$y$}
\includegraphics[width=\textwidth]{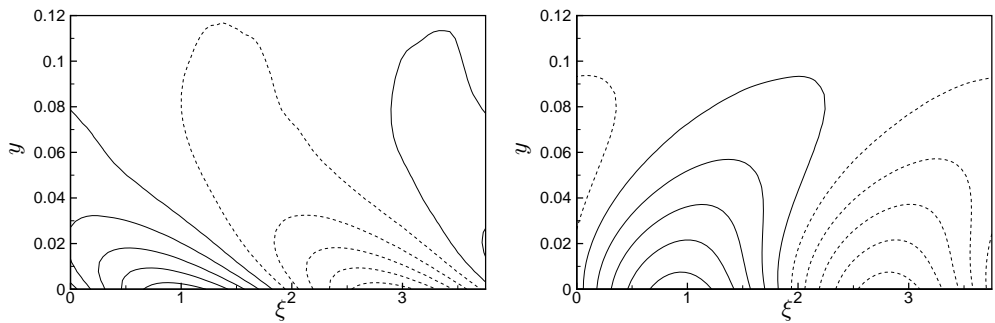}
\caption{Contour lines of the $\averzt{w}$ field. Contours are $(-0.45,0.1,0.45)$, dashed lines correspond to negative values. In this figure and in the following ones, the graphs are for one streamwise wavelength. The left graph refers to the DI case $(\omega=0.88,\kappa_x=1.66)$, while the right one refers to the DR case $(\omega=0.16,\kappa_x=1.66)$. For both cases $A=0.5$ and $Re=4760$.}
\label{fig:mean-w}
\end{figure}

\begin{figure}
\centering
\psfrag{x}{$\xi$}
\psfrag{y}{$y$}
\psfrag{a}{$\langle uu \rangle_{z,t}$}
\psfrag{b}{$\langle vv \rangle_{z,t}$}
\psfrag{c}{$\langle ww \rangle_{z,t}$}
\psfrag{d}{$\langle uv \rangle_{z,t}$}
\psfrag{e}{$\langle uw \rangle_{z,t}$}
\psfrag{f}{$\langle vw \rangle_{z,t}$}
\includegraphics[width=\textwidth]{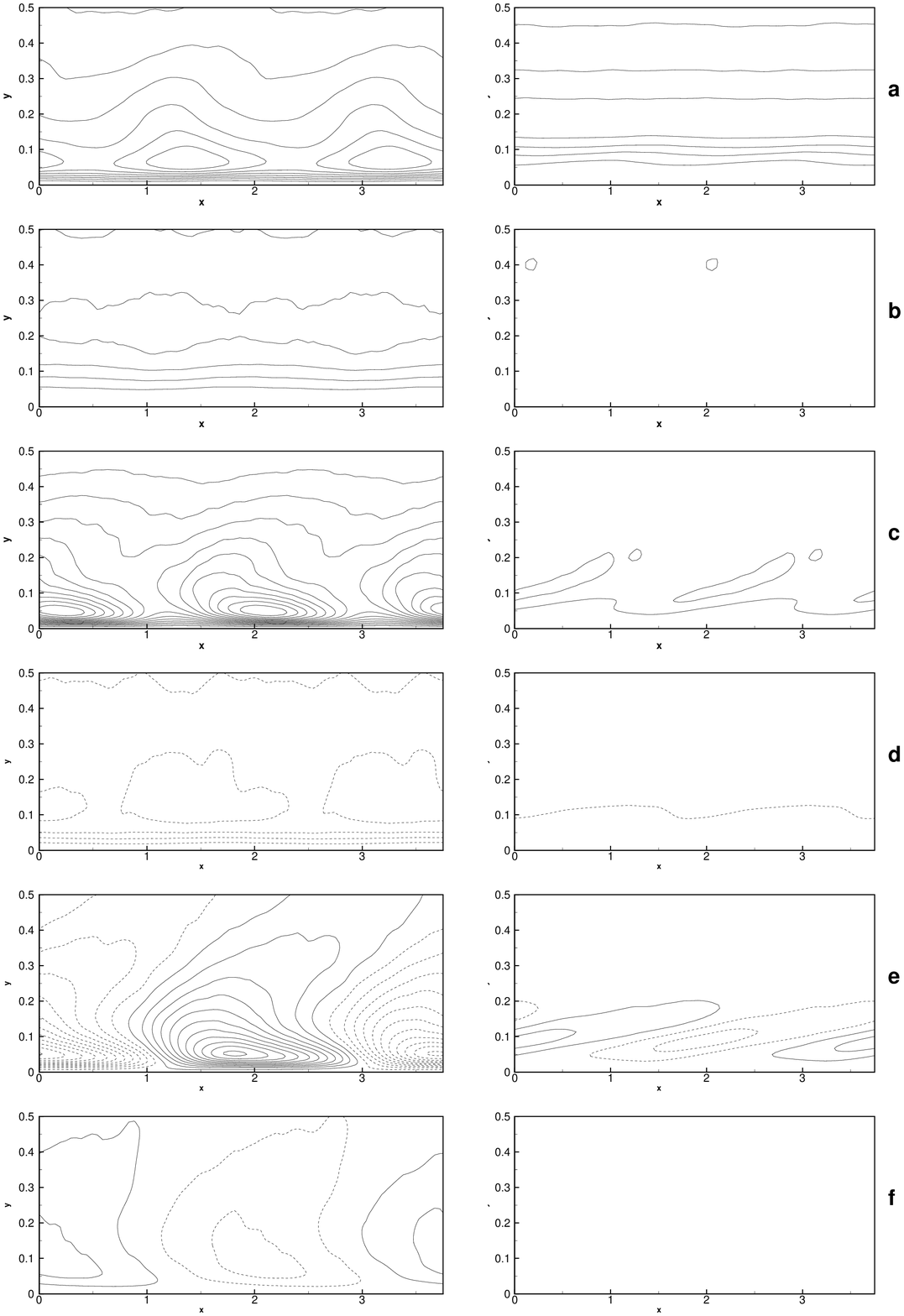}
\caption{Components of the Reynolds stress tensor. From top to bottom: $\averzt{uu}$, $\averzt{vv}$, $\averzt{ww}$, $\averzt{uv}$, $\averzt{uw}$ and $\averzt{vw}$. Contours are $(0,0.0005,0.01)$ for the diagonal components, and $(-0.0042,0.0004,0.0042)$ for the off-diagonal components; dashed lines correspond to negative levels. The graph on the left refers to the DI case $(\omega=0.88,\kappa_x=1.66)$, while the graph on the right refers to the DR case $(\omega=0.16,\kappa_x=1.66)$. For both cases $A=0.5$ and $Re=4760$.}
\label{fig:reystresses}
\end{figure}

The mean spanwise velocity and the components of the Reynolds stress tensor are now considered. The averaging is over time and $z$, at different oscillation phases over the convective variable $\xi = x - c t$, which is the proper phase alignment with the wall motion. (Hereinafter, this averaging operator is indicated by $\averzt{\cdot}$.) In figures \ref{fig:mean-w} and \ref{fig:reystresses}, the quantities are represented by contour plots as functions of $\xi$ and $y$. The streamwise direction is from left to right and for each case the left plot refers to the DI case, while the right contour indicates the DR case. The computational domain spans a streamwise distance of five wavelengths, but only quantities corresponding to one wavelength are shown.

Figure \ref{fig:mean-w} displays the mean spanwise velocity, i.e. $\averzt{w}$. Although the effect on the friction drag is opposite in the two cases, the appearance of the two fields is rather similar. A phase lag of spanwise velocity along the vertical direction is observed: at a non-zero wall-normal location, the value of the wall velocity is encountered at a different $\xi$ position. Similarly to the case of uniform oscillations, the convecting flow over the traveling waves is exposed to an unsteady transversal boundary layer. The thickness of the two $\xi$-modulated boundary layers in figure \ref{fig:mean-w} is comparable.

The six components of the Reynolds stress tensor, i.e. $\averzt{u_i u_j}$, are shown in figure \ref{fig:reystresses}. As expected, the intensity of all the components consistently increases in the DI case (when compared to the reference flow, where these quantities are a function of $y$ alone) and is significantly attenuated in the DR case. All the quantities in the DI case except $\averzt{vv}$ show a marked modulation along $\xi$, consistently with the instantaneous snapshot of figure \ref{fig:snapshot} (bottom). They can be qualitatively connected to the statistics of the reference flow, if the locally changing flow  direction is accounted for. The near-wall peaks of $\averzt{uu}$, for example, are related to similar peaks of $\averzt{ww}$, at different phases.

The variation along $\xi$ is less intense in the DR case. The reduction of the levels of $\averzt{uv}$, pertaining to the mean flow effects on the wall friction, is particularly noteworthy. The peak of the $\averzt{uu}$ contours is closer to the wall in the DI case than in the DR case, a well-known phenomenon in the oscillating-wall flow, for which the viscous and the buffer layer thicken when friction decreases \citep{choi-debisschop-clayton-1998}. For the $\averzt{vw}$ component, the level of fluctuations is so small in the DR case that no contour is shown on the chosen contour scale. Note moreover that $\averzt{uw}$ and $\averzt{vw}$ are zero in the reference flow.

\section{Discussion}
\label{sec:discussion}

The mechanism by which the traveling waves affect the turbulent friction is described in this section. We surmise that this mechanism includes the one of the oscillating wall, complemented by an additional effect responsible for drag increase and due to the phase speed attaining a finite value.

This additional effect is linked with the fact that the turbulent flow at the wall possesses the convective velocity scale $\sU_w=0.5$ \citep{kim-hussain-1993}. Indeed, when the phase speed of the waves approaches $\sU_w$, a large DI occurs. This may be interpreted by a phase lock between the near-wall flow structures and the wall waves, which is strongest along the line of maximum DI in figure \ref{fig:k-omega-DR}, i.e. when $c=\sU_w$. Since $\sU_w$ is a single velocity value computed from flow statistics, it should be viewed as representative of the (different) convection speeds of the near-wall structures, which can be smaller or larger than $\sU_w$. Consistently with this interpretation, the triangular region where DI occurs is identified by a range of phase speeds. When DI occurs, the structures are cyclically tilted by the wall forcing, which is stationary when observed in a reference frame convecting with the structures. The resultant streamwise modulation of the structures in the DI case is well illustrated in figure \ref{fig:snapshot}.

Away from the triangular region, the mechanism by which the traveling waves influence the friction drag becomes similar to that of the oscillating wall. In this latter case of uniform motion, the generally accepted qualitative explanation \citep{karniadakis-choi-2003} is that, when the wall oscillates in time with a period comparable with an optimal $T_{opt} = 2 \pi / \omega_{opt}$, the transversal Stokes layer effectively interacts with the near-wall turbulent structures, weakens the near-wall viscous cycle and produces a large DR. The optimal period is related to a time scale representing the typical life time of the most statistically significant turbulent structures \citep{quadrio-luchini-2003}, and to an optimal thickness of the Stokes layer. At periods smaller than $T_{opt}$, the Stokes layer is thin and the interaction with the turbulent flow is limited to the very near-wall region. DR increases with $T$ at these small periods. When $T$ is larger than $T_{opt}$, the Stokes layer becomes too thick and the interaction with the turbulent flow is not as effective: the DR effect rapidly deteriorates for increasing $T$.

This general picture can be easily adapted to the traveling waves. A first occurrence that points toward such a generalization is that the thickness of the transversal boundary layer assumes very similar values when the oscillating wall, the standing wave and the traveling wave guarantee the maximum DR. If the thickness is defined as the wall-normal location where the maximum (over all phases) of the spanwise velocity reduces to $\mathrm{e}^{-1} A$, the optimal thickness corresponding to maximum DR is $3.15 \cdot 10^{-2}$ for the oscillating wall, $3.6 \cdot 10^{-2}$ for the steady waves, and $3.4 \cdot 10^{-2}$ for the traveling waves.

It can therefore be conjectured that in all the three cases the distance of viscous diffusion from the wall toward the core of the turbulent flow is important in determining the turbulence suppression. The key difference with the oscillating wall is that the period to be compared with $T_{opt}$ is not the oscillation period $T$ of the traveling wave, but an equivalent period $\T$ which is related to the convective nature of the near-wall turbulent fluctuations, and to both $\omega$ and $\kappa_x$. We define $\T$ as the time required by the near-wall turbulent structures to cover a distance of one wavelength $\lambda_x$ along the streamwise direction at the relative velocity $\sU_w - c$. The simple formula for $\T$ is
\begin{equation}
\label{eq:teq}
\T \equiv \frac{\lambda_x}{\sU_w - c}.
\end{equation}
$\T$ is thus the time interval between two consecutive instants at which a near-wall structure convecting at velocity $\sU_w$ passes over the waves at the same phase.

Analogously to the oscillating wall, two cases can be distinguished. When $|\T|<T_{opt}$, or, equivalently, when
\begin{equation}
\label{eq:dr-unsteady-regime}
\left|\sU_w \kappa_x - \omega \right| > \omega_{opt},
\end{equation}
the forcing due to the traveling wave yields a thin transversal boundary layer, the bulk flow remains aligned along the streamwise direction, and DR  at constant $\kappa_x$ increases with $|\T|$. When $|\T| > T_{opt}$, i.e. when
\begin{equation}
\label{eq:dr-steady-regime}
\left|\sU_w \kappa_x - \omega \right| < \omega_{opt},
\end{equation}
DR  at constant $\kappa_x$ instead decreases with $|\T|$.

In the $\omega - \kappa_x$ plane, the points satisfying (\ref{eq:dr-unsteady-regime}) identify the exterior of the oblique strip bounded by two parallel straight lines with slope $\kappa_x / \omega = (\sU_w)^{-1}$ (and thus parallel to the line of maximum DI) passing through the points $\omega=\pm \omega_{opt},\kappa_x=0$.  The condition $\T = T_{opt}$ identifies the maxima at $\omega=0,\kappa_x=\kappa_{x,opt}=1$ for the standing waves, and at $\kappa_x=0,\omega=\omega_{opt}=-0.5$ for the oscillating wall. For backward traveling waves, the analogy with the oscillating-wall regime holds. Moving along lines of constant $\kappa_x$, DR increases with $\T$, except, as expected, inside the strip where DR decreases with $\T$. The red ridge of maximum DR in figure \ref{fig:k-omega-DR} and the border of the strip $\T=T_{opt}$ show good agreement.

For forward traveling waves, the behaviour of DR is determined by the mutual action of the two mechanisms outlined above. The ridge of maximum DR now lies entirely at the exterior of the strip and not at the border of it. This is because the DI region widens at large $\kappa_x$ and dominates the oscillating-wall condition to dictate the changes in drag. 
Points laying outside the strip {\em and} outside the influence of the triangular region show increasing DR at fixed $\kappa_x$ for increasing $|\T|$. At small $\kappa_x$, points at low frequency are inside the strip where DR drops as $|\omega|$ decreases, ultimately falling into the infrared DI region. 

\begin{figure}
\vspace{1cm}
\centering
\psfrag{o}{$\omega$}
\psfrag{k}{$\kappa_x$}
\psfrag{M}{Curve of maximum DR}
\psfrag{oscillating}{(oscillating wall)}
\psfrag{steady}{(standing waves)}
\psfrag{MDI}{Line of maximum DI}
\psfrag{DI}{Region of DI}
\psfrag{Te}{$|\T| = T_{opt}$}
\psfrag{MDR}{Maximum DR}
\psfrag{traveling}{(traveling waves)}
\vspace{0.5cm}
\includegraphics[width=\textwidth]{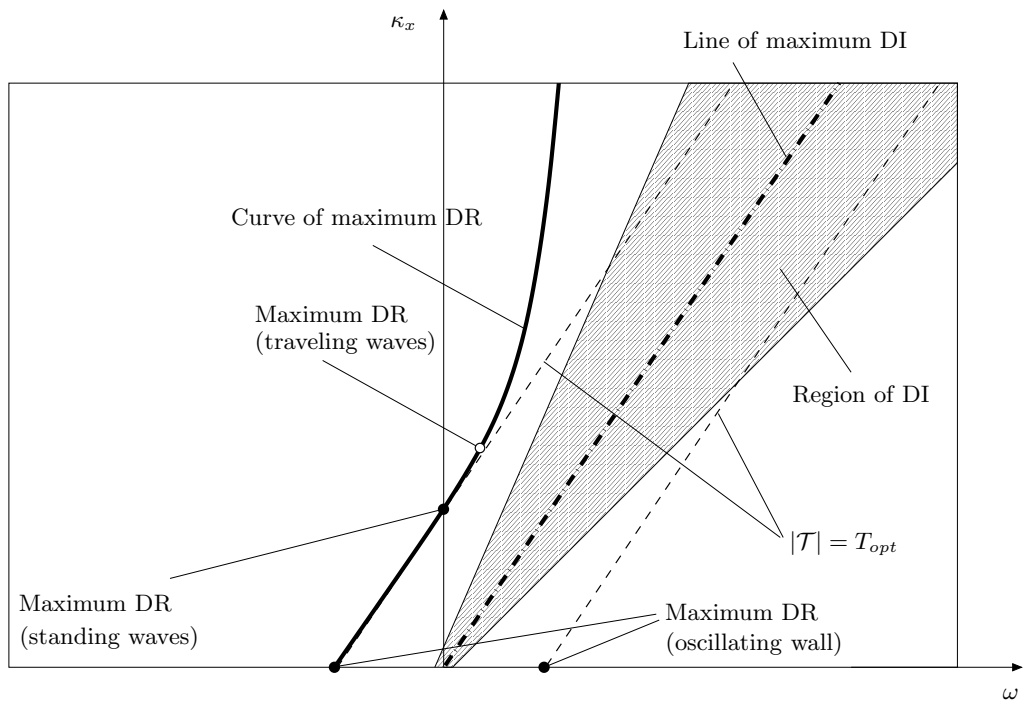}
\caption{Qualitative sketch of the map of DR due to the traveling waves in the $\omega - \kappa_x$ plane. DR occurs in the entire plane, except in the dashed area where DI takes place. Thick lines indicate loci of maximum DR (solid) and maximum DI (dash-dotted) at given $\kappa_x$. The region confined between the two oblique dashed lines is where the wall forcing is slower than the optimal timescale, as observed in the moving reference frame (see text). Symbols indicate points of maximum DR on the axes (closed circles) and on the entire plane (open circle).}
\label{fig:flowregimes}
\end{figure}

The sketch in Figure \ref{fig:flowregimes} summarizes the effects of the traveling waves on the turbulent flow. The thick solid curve denotes the locus of points of maximum DR at given $\omega$. The friction drag decreases everywhere in the plane, except in the cone-shaped area, where DI occurs. The dash-dotted line indicates the points of maximum DI at given $\kappa_x$, occurring when $c = \sU_w$, or, equivalently, when $\T \rightarrow \infty$. The dashed lines mark the strip defined by $|\T| = T_{opt}$.

\subsection{Open questions}

Several interesting questions are left unanswered by the discussion above. A non-exhaustive list includes: Can the details of the DR mechanism be better understood? What are the real possibilities of this technique in terms of net power savings? Do particular waveforms exist, possibly different from sinusoids, that allow for further optimizations? There is clearly need for a further understanding of the mechanisms by which the wall friction is modified by the traveling waves.

An additional, important issue is how to exploit the present results for practical purposes. This is unfortunately not immediate, but a sound understanding of the complex flow physics is definitely an essential preliminary step. As discussed by \cite{viotti-quadrio-luchini-2008}, the discovery of a preferred longitudinal length scale selected by the standing waves could have a significant impact on the design of optimal wall roughness for drag reduction. Recent works have addressed the application of similar traveling-waves techniques. \cite{zhao-wu-luo-2004} have pointed out that an array of smart materials and MEMS would render a smart flexible wall the most promising candidate for effective turbulence control. \cite{itoh-etal-2006} have succeeded in studying experimentally the spanwise-traveling wave concept in a turbulent boundary layer through a flexible sheet.

One further point of interest is the prediction of the drag modification (without resorting to direct measurements either via full DNS or laboratory experiments). In the oscillating-wall case, the space-averaged velocity profile follows closely the laminar solution, which has led to a scaling parameter for drag reduction \citep{choi-xu-sung-2002, quadrio-ricco-2004}. Analogously, \cite{viotti-quadrio-luchini-2008} have found that the laminar layer induced by the standing waves agrees well with the space-averaged turbulent profile. Future work should therefore be directed at extending these results to the traveling-wave case to estimate the modification of drag.

\section{Summary}
\label{sec:summary}

This work has described the response of a turbulent channel flow to sinusoidal waves of spanwise velocity applied at the wall and traveling along the streamwise direction. The effects of the wavenumber $\kappa_x$, temporal frequency $\omega$ and -- to a lesser extent -- forcing amplitude $A$ and Reynolds number $Re$ have been explored via Direct Numerical Simulations. Such a demanding parametric study, that required centuries of CPU time, would have been impossible, had our special computing system not been available.

In the $\omega - \kappa_x$ plane, a region where drag reduction (DR) is very large has been identified. This region pertains to forward-traveling waves with small phase speed, but extends in the region of backward-traveling waves at small $\kappa_x$. A maximum DR of 48\% is measured for $Re=4760$ and $A=0.5 U_P$; DR above 45\% is observed over an extremely wide range of wavelengths, provided the frequency is adjusted to maintain the correct phase speed. DR weakly decreases with the Reynolds number, going from full relaminarization of the flow at $Re= 2175$ to 42\% DR at $Re=10500$. DR has been found to increase with $A$ (we have measured up to 60\% for $A=1.2 U_P$). The global energy budget has also been addressed: the energetic cost for producing the traveling waves is nearly minimum where DR is maximum, which implies large potential net savings.

Backward-traveling waves always yield DR, while forward-traveling waves may also result in significant drag increase (DI) at finite phase speeds. Small levels of DI have also been observed at very low frequencies (infrared DI) for the case of uniform spanwise oscillations ($\kappa_x=0$).

Flow visualizations and turbulence statistics have been useful to show that, when DI takes place, the flow structures are tilted significantly along the spanwise direction and present a marked streamwise modulation, whereas they remain aligned with the streamwise direction when DR occurs.

DI is observed when the waves travel with a phase speed comparable to the convection velocity of near-wall turbulent fluctuations. The near-wall structures are cyclically tilted by forcing, which ultimately results in an increased drag. At different phase speeds, DR is produced with a mechanism analogous to that of the oscillating wall. The optimal value of the oscillation period for the oscillating-wall regime still dictates the optimal conditions for the traveling waves, but the relevant period is that observed in a reference frame that convects with the flow.

\section*{Acknowledgments}
The standing collaboration with Professor P. Luchini and the use of the dedicated parallel computing system at the Universit\`a di Salerno are gratefully acknowledged. We have benefited from discussions with Mr F. Martinelli. This work was partially supported from the Italian Ministry of University and Research through the grant PRIN 2005 on {\em Large scale structures in wall turbulence}.

\bibliographystyle{jfm}
\bibliography{../mq}

\end{document}